\documentclass[conference]{IEEEtran} 

\usepackage{graphicx, amsmath,mathrsfs, multirow, array,  epstopdf, subfigure, amssymb,  cite, fixmath, dsfont, mathtools,arydshln}

\usepackage[short]{optidef}
\usepackage[utf8x]{inputenc}
\usepackage[binary-units]{siunitx}

\usepackage{url}

\begin{document}
\title{Reconfigurable Intelligent Surface-assisted Edge Computing to Minimize Delay in Task Offloading}

\author{\IEEEauthorblockN{Mithun Mukherjee\IEEEauthorrefmark{1}, Vikas Kumar\IEEEauthorrefmark{2}, Suman Kumar\IEEEauthorrefmark{3}, 
Jaime Lloret\IEEEauthorrefmark{4},
Qi Zhang\IEEEauthorrefmark{5}, 
and Mian Guo\IEEEauthorrefmark{6}}

\IEEEauthorblockA{\IEEEauthorrefmark{1}School of Artificial Intelligence, Nanjing University of Information Science and Technology, China, m.mukherjee@ieee.org}

\IEEEauthorblockA{\IEEEauthorrefmark{2}Bharat Sanchar Nigam Limited, India, vikas.kr@bsnl.co.in}

\IEEEauthorblockA{\IEEEauthorrefmark{3}Department of Mathematics, IGNTU Amarkantak, MP, India, suman@igntu.ac.in}

\IEEEauthorblockA{\IEEEauthorrefmark{4}Universitat Politecnica de Valencia, Spain, jlloret@dcom.upv.es}

\IEEEauthorblockA{\IEEEauthorrefmark{5}DIGIT,  Department of Electrical and Computer Engineering, Aarhus University,   Aarhus, Denmark, qz@ece.au.dk}

\IEEEauthorblockA{\IEEEauthorrefmark{6}School of Electronics and Information, Guangdong Polytechnic Normal University, P.R. China, mian.guo@ieee.org }
}
\maketitle
\begin{abstract}
The advantage of computational resources in edge computing near the data source has kindled growing interest in delay-sensitive Internet of Things (IoT) applications. However, the benefit of the edge server is limited by the uploading and downloading links between end-users and  edge servers when these end-users seek computational resources from edge servers. The scenario becomes more severe when the user-end's devices are in the shaded region resulting in low uplink/downlink quality. In this paper, we consider a reconfigurable intelligent surface (RIS)-assisted edge computing system, where the benefits of RIS are exploited to improve the uploading transmission rate. We further aim to minimize the delay of worst-case in the network when the end-users either compute task data in their local CPU or offload task data to the edge server. Next, we optimize the uploading bandwidth allocation for every end-user's task data to minimize the maximum delay in the network. The above optimization problem is formulated as quadratically constrained quadratic programming. Afterward, we solve this problem by semidefinite relaxation. Finally, the simulation results demonstrate that the proposed strategy is scalable under various network settings. 
\end{abstract}

\section{Introduction}
In recent years, several industries have been focusing their technological advancement towards high performance computing in cloud data centers. For example, in 2020, NVDIA announced the potential use of DGX A100 NVIDIA’s third-generation Artificial Intelligence (AI) system box~\cite{NVIDIADGX-2} that is aimed at the massive gain in performance for AI-related and cutting-edge applications with less power consumption. At the same time, we are witnessing the paradigm changing from constituting a well-run centralized data center infrastructure to the network edge~\cite{Liu2019MEC,ZhangIoT2021, MithunTactile,JaimeMag2018}, particularly when there is a need to deliver proximity, low-latency, and reliable services for the mission-critical applications, such as remote-surgery, industrial automation and driverless cars. The leading industries with their cloud service providers (e.g., EGX Edge AI platform, NVIDIA RTX graphics with CloudXR, GPU virtualization, and Qualcomm Technologies’ Boundless XR client optimizations~\cite{QualcommXR}
and EdgeConneX~\cite{EdgeConnecX}) are making their way for the deployment of edge-assisted service provisioning.

\subsection{Motivation}
Although MEC brings computational, caching, and storage resources towards the network edge, the connectivity and coverage of the access points and base station play critical roles. To say, when end-users aim to avail the computing, caching, and storage resources of edge server, they need to rely on the wireless channels.  
Basically, irrespective of application type and service, the uploading/downloading time is an important factor in delay-sensitive service provisioning. This becomes even worse when the network coverage is poor near the cell edge or blocked by obstacles. The uploading and downloading rate and the resulting latency are significantly affected by communication resource allocation. This, in turn, affects the computation delay of end-user's task. To address the above shortcoming that arises due to the connectivity and coverage of the network services, reconfigurable intelligent surface (RIS)~\cite{BasarRIS2019} can assist MEC. 

\subsection{Related Work}
The role of RIS has been studied in the MEC system, where the end-users aim to offload their computation-intensive tasks to the edge server that resides at the access point~\cite{Bai_Nallanathan_JSAC2020, Bai_Nallanathan_TWC2021}. They formulated a latency minimization problem by optimizing the task offloading data size, edge computing resource allocation and RIS phase shift coefficients. To maximize the total amount of data (in terms of bits) processed by end-users and edge server, Chu et al.~\cite{Chu_Shojafar_WCL2021} suggested how to adjust the phase shift of the RIS in addition to the transmit power and time allocation for the end-users, and edge server's computing resource allocation for the end-users. Another study in~\cite{liu_Dusit_2020} shows how the edge server adjusts the RIS controller to maximize its revenue while guaranteeing the customized information rate for each end-user. Later, another parallel work studied the RIS-enabled MEC system in~\cite{Zhou_RuiZhang_WCL2021}. Again, this was to minimize the latency, which is basically calculated as the sum of two end-user's computation offloading time. Moreover, Cao et al.~\cite{Cao_mmWave_2021} have shown how RIS can resolve the link blockage problem in the mm-Wave MEC system to guarantee real-time offloading from the end-users. This is an interesting and detailed study on how RIS can directly affect the task offloading chances for the end-users that suffer from mm-Wave link blockage. 
 
Recently, over-the-air computation (AirComp)~\cite{LiuAiRComP2020} that integrates communication and computation has attracted academia and industries' attention due to its fast data aggregation from IoT devices. However, due to the unreliable channel conditions, the performance of AirComp is severely limited. To address this, RIS~\cite{IRSWework13,ZhibinAirCompRIS, mithunarxiv2021} has been found a suitable candidate to assist the uplink and downlink transmission.

\subsection{Our Contributions and Organization}
We summarize our main contributions as follows: We consider an RIS-assisted edge computing system, where end-users offload their task data to the edge server to minimize the overall delay. We aim to leverage the benefits of RIS for the uplink transmission rate in data offloading to the edge server. With the assistance of RIS, a delay-minimization problem is formulated by optimizing the offloading decision variables and bandwidth allocation for the offloaded task data. We formulate the above optimization problem as Quadratically Constrained Quadratic Programming (QCQP) problem. Afterward, we apply semi-definite relaxation (SDR) to solve the problem. Finally, we show that the proposed offloading strategy with RIS can achieve better performance than without RIS assistance and \emph{local CPU only} approaches.

The rest of the paper is organized as follows. In Section~\ref{sec:SystemModel}, we discuss the RIS-assisted MEC system model. We formulate the optimization problem and apply SDR in Section~\ref{sec:ProblemFormulation}. The simulation results are presented in Section~\ref{sec:SimulationResult}. Finally, we conclude our work in Section~\ref{sec:Conclusions}.

\section{System Model}\label{sec:SystemModel}

\begin{figure}[t]
\centering
\includegraphics[width={0.44\textwidth}]{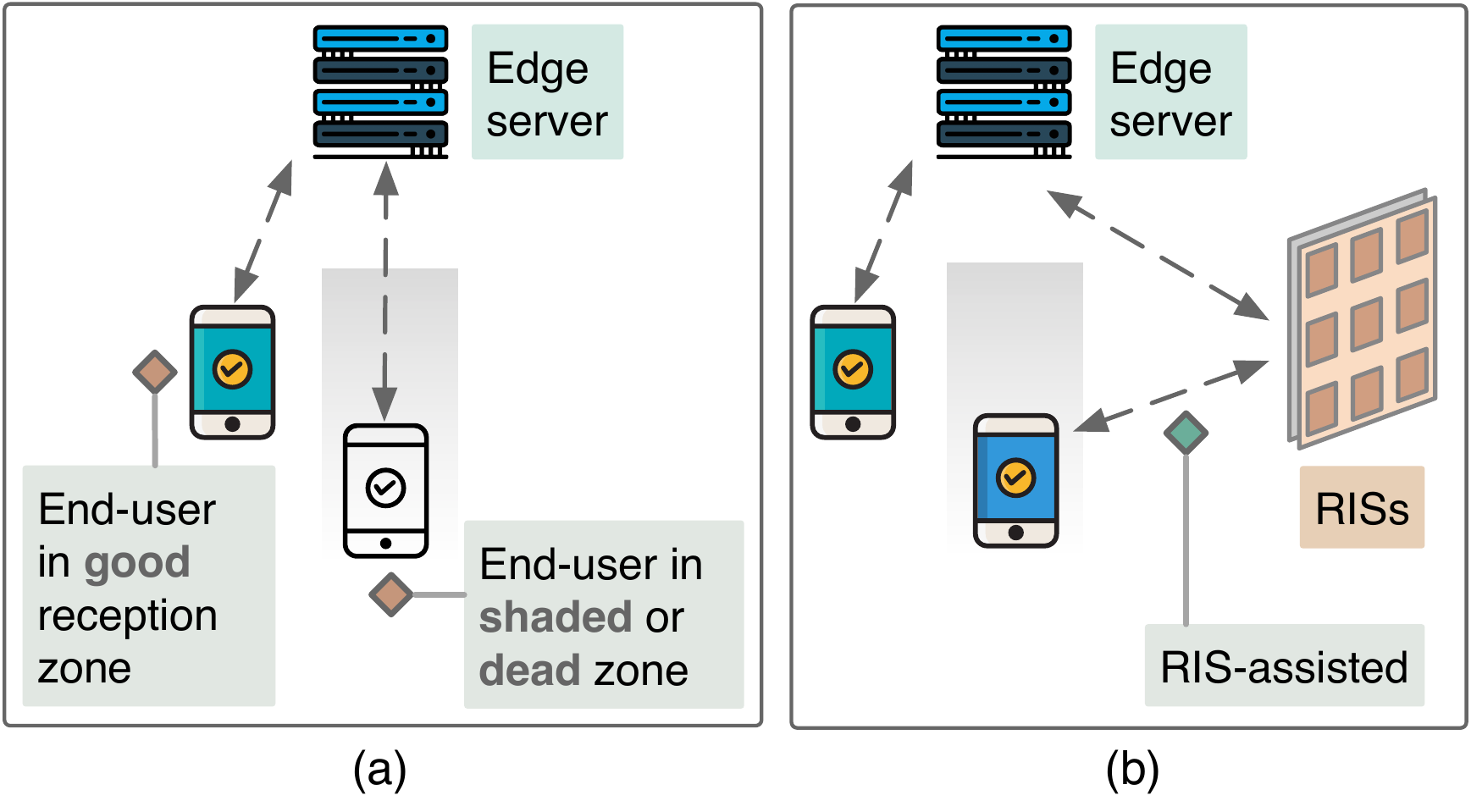}
\caption{(a) An illustration of a RIS-assisted edge computing system with end-users under \emph{good} and \emph{shadowed} region. (b) The RIS can assist to improve the uplink and downlink quality for the end-users under shadowed region.}
\label{fig:SystemModel}
\end{figure} 

We consider an RIS-assisted edge computing system, as shown in Fig.~\ref{fig:SystemModel}. The set of the end-users is denoted as $\mathcal{M} = \{1,2,\ldots, M\}$, where $M$ is the total number of end-users in the network. Due to the limited computational resources (in terms of CPU speed) in local CPU, these end-users often offload task data to edge server when the tasks demand fast processing. Note that among these end-users, we assume that $K$ end-users have good reception quality and the remaining $(M-K)$ end-users are in poor signal reception area. 
We denote the end-users in \emph{good} and \emph{poor} signal reception areas as the $i$th and the $j$th end-user, respectively, where $i= 1,\,\ldots, \,K$ and $j = (K+1),\, \ldots, \, M$.
We further denote the offloading decision variable for the $m$th end-user at local device of each end-user as $x_{m}$ and  the task processing decision variable at edge server for the end-user as $y_{m}$, where $m = 1,\, \ldots, \, M$ and 
\begin{align*}
x_m=& \begin{cases}
\multirow{2}{*}{1}& \text{when the $m$th end-user's task data is }\\ 
& \text{locally processed,}\\
\multirow{1}{*}{0}& \text{otherwise,}\\
\end{cases}\\
y_m=& \begin{cases}
\multirow{2}{*}{1}& \text{when the $m$th end-user's task data is  }\\ 
& \text{offloaded and processed at the edge server,}\\
\multirow{1}{*}{0}& \text{otherwise.}\\
\end{cases}
\end{align*}
Note that $x_{m}+y_{m} = 1$. Moreover, these binary decision variables satisfy $x_{m}(1-x_{m}) = 0$  and
$y_{m}(1-y_{m}) = 0$. 

\subsection{Local Computing Delay}
When a task is locally processed by the end-user, the computation  delay becomes
\begin{subequations}
\begin{align}
T^L_i & = \dfrac{x_{i}\, D_i\,L}{f_i^l}\:[\text{s}] \quad i\in \{1,2,\ldots, K\},\\
T^L_{j} & = \dfrac{x_{j}\, D_{j}\,L}{f_{j}^l}\:[\text{s}]\quad {j}\in \{(K+1),\ldots, M\}, 
\end{align}
\end{subequations}
where $D_i\text{ and }D_{j}$ is the input data size [bits] of the $i$th and $j$th end-user, respectively, $L$ is the processing density [CPU cycles/bit] for a task, and $f_i$ and $f_j$ denote the CPU clock speed [CPU cycles/s] of the $i$th and $j$th end-user, respectively. We assume equal task processing density for every end-users. 

\subsection{Offloading Delay for End-users without RIS Assistance}
Basically, this is the case when  the end-users are in \emph{good} signal reception area.  When the $i$th end-user task data is offloaded and processed at the edge server, the offloading delay becomes
\begin{align}
T^\textsf{E}_i = y_i\bigg(\underbrace{\dfrac{D_i}{\eta_i \,\beta_{i}\,C}}_{\text{uploading}}+ \underbrace{\dfrac{D_i\,L}{f_i^e}}_{\text{computation}}\bigg)\:[\text{s}]\:,
\end{align}
where $\eta_i$ is the spectral efficiency of uplink transmission between the $i$th end-user and edge server, $C$ is the total uplink bandwidth, $\beta_i$ is the fraction of total uplink bandwidth allocated to the $i$th end-user, and $f_i^e$ is the CPU rate allocated by the edge server to process the $i$th end-user's offloaded task.
\subsection{Offloading Delay for End-users with RIS Assistance}
When the $j$th end-user with poor wireless connection offloads its task data to the edge server with RIS assistance, the offloading delay can be written as
\begin{align}
T^\textsf{E}_{j} = y_{j}\bigg(\underbrace{ \dfrac{D_{j}}{\eta_{j}\,\beta_{j}\,C}}_{\text{uploading}}+ \underbrace{\dfrac{D_{j}\,L}{f_{j}^e}}_{\text{computation}}\bigg)\:[\text{s}] \:,
\end{align}
where $\eta_j$ is the spectral efficiency of RIS-assisted uplink transmission between the $j$th end-user and the edge server, $\beta_{j}$ is the fractional value of total uplink bandwidth allocated to the $j$th end-user and $f_j^e$ is the CPU rate  allocated by edge server to process the $j$th end-user's offloaded task.

\section{Problem Formulation}\label{sec:ProblemFormulation}
We write the delay of the worst case in the network as $\max \{(T_{i}^\textsf{L}+ T_{i}^\textsf{E}),\,(T_{j}^\textsf{L}+T_{j}^\textsf{E})\}\forall i\in\mathcal{N}_1,\, j\in\mathcal{N}_2$, where $\mathcal{N}_1=\{1\ldots K\}$, $\mathcal{N}_2=\{K+1,\ldots M\}$. 
We aim to minimize the maximum delay by jointly optimizing the task offloading decision vector $\boldsymbol{\xi} = {[x_{m},  y_{m}]}^{\intercal}$ and the bandwidth allocation vector $\mathbf{r} = [\beta_{m}]^{\intercal}$, where
\begin{subequations}\label{UEdelay}
\begin{align}
T^\textsf{L}_i+T^\textsf{E}_i & = \dfrac{D_i\,L\,x_{i}}{f_i^l}
 +\dfrac{D_i\,y_{i}}{\eta_i\,\beta_i\,C} +  \dfrac{D_i\,L\,y_{i}}{f_i^e}\:[\text{s}] \:,\\
T^\textsf{L}_{j}+T^\textsf{E}_{j} & = \dfrac{D_{j}L x_{j}}{f_{j}^l}
 +\dfrac{D_{j} y_{j}}{\eta_{j}\,\beta_{j}\,C} + \dfrac{D_{j} L y_{j}}{f_{j}^e}\:[\text{s}] \:.
\end{align}\label{1stcvx0}
\end{subequations}
We define the above optimization problem as
\begin{mini!}[2]
{\boldsymbol{\xi},\mathbf{r}}{\max \{(T_{i}^\textsf{L}+ T_{i}^\textsf{E}),\,(T_{j}^\textsf{L}+T_{j}^\textsf{E})\}  \; \forall ~ i\in\mathcal{N}_1,\, j\in\mathcal{N}_2} {}{}{\label{eq:pb1_0}}
\addConstraint{x_{m}(1-x_{m})=0}{}
\addConstraint{y_{m}(1-y_{m})=0}{}
\addConstraint{x_{m}+y_{m}=1}{}
\addConstraint{\sum\limits_{i = 1}^{K} {\beta_i} +\sum\limits_{{j} = K+1}^{M} {\beta_{j}} \le 1}{\label{eq:pb0_1}}{},
\end{mini!}
where the constraint \eqref{eq:pb0_1} corresponds to the total uplink bandwidth $C$. Now, we take an auxiliary variable $t$ as
\begin{align}
    \max\limits_{i\in\mathcal{N}_1,{j}\in\mathcal{N}_2} \big\{(T_{i}^\textsf{L}+ T_{i}^\textsf{E}),\,(T_{j}^\textsf{L}+T_{j}^\textsf{E})\big\} = t\:,\label{1stcvx1}
\end{align}
then, from \eqref{1stcvx0} and \eqref{1stcvx1}, we write
\begin{subequations}
\begin{align}
 & \dfrac{D_i\,L\,x_{i}\,\beta_i}{f_i^l}
 +\dfrac{D_i\,y_{i}}{\eta_i\,C} +  \dfrac{D_i\,L\,y_{i}\,\beta_i}{f_i^e} -\beta_i\,t\le 0,\\
& \dfrac{D_{j}L x_{j}\beta_{j}}{f_{j}^l}
 +\dfrac{D_{j} y_{j}}{\eta_{j}\,C} + \dfrac{D_{j} L y_{j}\beta_{j}}{f_{j}^e}-\beta_{j}\,t \le 0.
 \label{eqn-aux-var-t2}
\end{align}
\end{subequations}
Accordingly, the optimization problem becomes
\begin{mini!}[2]
{\boldsymbol{\xi},\mathbf{r}}{t} {}{}{\label{eq:pb11_0}}
\addConstraint{x_{m}(1-x_{m})=0}{}
\addConstraint{y_{m}(1-y_{m})=0}{}
\addConstraint{x_{m}+y_{m}=1}{}
\addConstraint{\sum\limits_{i = 1}^{K} {\beta_i} +\sum\limits_{{j} = K+1}^{M} {\beta_{j}} \le 1}
\addConstraint{\text{(7a) and (7b)}\:.}{\label{eq:pb11_1}}{}
\end{mini!}

\subsection{Vector-matrix Formation}
Now, we denote $\mathbf{w}\!=\![x_1,x_2,\ldots, x_K,x_{K+1},\ldots, x_M,y_1,$ $y_2,\ldots, y_K,y_{K+1},\ldots, y_M,\!\beta_1,\!\beta_2,\!\ldots, \beta_K,\beta_{K+1},\ldots, \beta_M,\, t]^{\intercal}$ and define the unit vector as ${\mathbf{e}}_q = [\mathbf{0}_{1 \times (q-1)},\,1,$ $\mathbf{0}_{1 \times (3M+1-q)}]^{\intercal}$. Then, the matrix form of problem (8) can be expressed as

\begin{mini!}[2]
{{\mathbf{w}}}{{\mathbf{e}}^{\intercal}_{(3M+1)}{\mathbf{w}}} {}{}{\label{eq:pb12_0}}
\addConstraint{{\mathbf{w}}^\intercal \mathbf{A}_{x,m} {\mathbf{w}} - {\mathbf{e}}^{\intercal}_{m} {\mathbf{w}}=0}{\label{eq:pb12}}{}
\addConstraint{{\mathbf{w}}^\intercal \mathbf{A}_{y,m} {\mathbf{w}} - {\mathbf{e}}^{\intercal}_{m+M} {\mathbf{w}}=0}{\label{eq:pb12}}{}
\addConstraint{{\mathbf{e}}^{\intercal}_{m} {\mathbf{w}}+ {\mathbf{e}}^{\intercal}_{m+M} {\mathbf{w}}=1}{\label{eq:pb12}}{}
\addConstraint{\sum\limits_{i = 1}^{K} {{\mathbf{e}}^\intercal_{i+2M} {\mathbf{w}}} \:+\sum\limits_{{j} = K+1}^{M} {{\mathbf{e}}^\intercal_{j+2M} {\mathbf{w}}} \le 1}{\label{eq:pb12_1}}{}
\addConstraint{{\mathbf{w}}^\intercal \mathbf{A}_{\beta x,i} {\mathbf{w}} +{\mathbf{w}}^\intercal \mathbf{A}_{\beta y,i} {\mathbf{w}} + {\mathbf{b}}_{cy,i}^\intercal {\mathbf{w}}\nonumber  \breakObjective{ \quad\quad\quad\quad\quad\quad\quad\quad+{\mathbf{w}}^\intercal \mathbf{A}_{\beta t,i} {\mathbf{w}}\le 0}}
\addConstraint{{\mathbf{w}}^\intercal \mathbf{A}_{\beta x,j} {\mathbf{w}} +{\mathbf{w}}^\intercal \mathbf{A}_{\beta y,j} {\mathbf{w}} + {\mathbf{b}}_{cy,j}^\intercal {\mathbf{w}} \nonumber \breakObjective{\quad\quad\quad\quad\quad\quad\quad\quad+{\mathbf{w}}^\intercal \mathbf{A}_{\beta t,j} {\mathbf{w}}\le 0\, ,}{}{\label{eq:pb12_9}}}
\end{mini!}

where
\begin{align}
& \mathbf{A}_{x,m} = \left[ {\begin{array}{c}
{\mathbf{0}_{(m-1) \times (3M+1)}}\\
\hdashline[2pt/2pt]
\mathbf{e}_{m}^{\intercal} \\
\hdashline[2pt/2pt]
{\mathbf{0}_{(3M+1-m) \times (3M+1)}}
\end{array}} \right], \nonumber \\
& \mathbf{A}_{y,m} = \left[ {\begin{array}{c}
{\mathbf{0}_{(M-1+m) \times (3M+1)}}\\
\hdashline[2pt/2pt]
\mathbf{e}_{(m+M)}^{\intercal} \\
\hdashline[2pt/2pt]
{\mathbf{0}_{(2M+1-m) \times (3M+1)}}
\end{array}} \right], \nonumber \\
& \mathbf{b}_{cy,i} = k_i^c{\mathbf{e}}_{M+i},\, 
 k_i^l= \dfrac{D_i\,L}{f_i^l},\,
 k_i^e= \dfrac{D_i\,L}{f_i^e},\, k_i^c= \dfrac{D_i}{\eta_i C},\nonumber \\
& \mathbf{A}_{\beta x,i}=\dfrac{k_i^l}{2}\left[ {\begin{array}{*{20}{c}} \mathbf{0}_{(i-1) \times (3M+1)}\\
 \hdashline[2pt/2pt]
 \mathbf{e}_{i+2M}^{\intercal} \\
 \hdashline[2pt/2pt]
 \mathbf{0}_{(2M-1) \times (3M+1)}\\ 
 \hdashline[2pt/2pt]
 \mathbf{e}_{i}^{\intercal} \\
 \hdashline[2pt/2pt]
 \mathbf{0}_{(M+1-i) \times (3M+1)}
 \end{array}} \right],\nonumber \\
& \mathbf{A}_{\beta y,i}=\dfrac{k_i^e}{2}\left[ {\begin{array}{*{20}{c}} \mathbf{0}_{(M-1+i) \times (3M+1)}\\
 \hdashline[2pt/2pt]
 \mathbf{e}_{i+2M}^{\intercal} \\
 \hdashline[2pt/2pt]
 \mathbf{0}_{(M-1) \times (3M+1)}\\ 
 \hdashline[2pt/2pt]
 \mathbf{e}_{i+M}^{\intercal} \\
 \hdashline[2pt/2pt]
 \mathbf{0}_{(M+1-i) \times (3M+1)}
 \end{array}} \right],\nonumber \\
 & \mathbf{b}_{cy,j} = k_{j}^c{\mathbf{e}}_{M+{j}},\, 
 k_{j}^l= \dfrac{D_{j}\,L}{f_{j}^l},\,
  k_{j}^e= \dfrac{D_{j}\,L}{f_{j}^e},\, k_{j}^c= \dfrac{D_{j}}{\eta_{j} C},\nonumber \\
& \mathbf{A}_{\beta t,i}=-\dfrac{1}{2}\left[ {\begin{array}{*{20}{c}} \mathbf{0}_{(2M-1+i) \times (3M+1)}\\
 \hdashline[2pt/2pt]
 \mathbf{e}_{3M+1}^{\intercal} \\
 \hdashline[2pt/2pt]
 \mathbf{0}_{(M-i) \times (3M+1)}\\ 
 \hdashline[2pt/2pt]
 \mathbf{e}_{i+2M}^{\intercal} 
 \end{array}} \right],\nonumber\\
 & \mathbf{A}_{\beta x,j}=\dfrac{k_{j}^l}{2}\left[ {\begin{array}{*{20}{c}} \mathbf{0}_{({j}-1) \times (3M+1)}\\
 \hdashline[2pt/2pt]
 \mathbf{e}_{{j}+2M}^{\intercal} \\
 \hdashline[2pt/2pt]
 \mathbf{0}_{(2M-1) \times (3M+1)}\\ 
 \hdashline[2pt/2pt]
 \mathbf{e}_{j}^{\intercal} \\
 \hdashline[2pt/2pt]
 \mathbf{0}_{(M+1-{j}) \times (3M+1)}
 \end{array}} \right],\nonumber \\
& \mathbf{A}_{\beta y,j}=\dfrac{k_{j}^e}{2}\left[ {\begin{array}{*{20}{c}} \mathbf{0}_{(M-1+{j}) \times (3M+1)}\\
 \hdashline[2pt/2pt]
 \mathbf{e}_{{j}+2M}^{\intercal} \\
 \hdashline[2pt/2pt]
 \mathbf{0}_{(M-1) \times (3M+1)}\\ 
 \hdashline[2pt/2pt]
 \mathbf{e}_{{j}+M}^{\intercal} \\
 \hdashline[2pt/2pt]
 \mathbf{0}_{(M+1-{j}) \times (3M+1)}
 \end{array}} \right],\nonumber 
     \end{align}
 \begin{align}
& \mathbf{A}_{\beta t,{j}}=-\dfrac{1}{2}\left[ {\begin{array}{*{20}{c}} \mathbf{0}_{(2M-1+{j}) \times (3M+1)}\\
 \hdashline[2pt/2pt]
 \mathbf{e}_{3M+1}^{\intercal} \\
 \hdashline[2pt/2pt]
 \mathbf{0}_{(M-j) \times (3M+1)}\\ 
 \hdashline[2pt/2pt]
 \mathbf{e}_{{j}+2M}^{\intercal} 
 \end{array}} \right]. \nonumber
  \end{align}
\subsection{QCQP Formulation}
 Defining ${\mathbf{z}}=[{\mathbf{w}}^{\intercal}\;1]^{\intercal}$, the problem (9) can be transformed into homogeneous separable QCQP formulation as follows
\begin{mini!}[2]
{{\mathbf{z}}}{{\mathbf{z}}^{\intercal}\, \mathbf{B}\,{\mathbf{z}}} {}{}{\label{eq:pb12_0}}
\addConstraint{{\mathbf{z}}^{\intercal}\, \mathbf{B}_{x,m}\,{\mathbf{z}} =0}{}{}
\addConstraint{{\mathbf{z}}^{\intercal}\, \mathbf{B}_{y,m}\,{\mathbf{z}} =0}{}{}
\addConstraint{{\mathbf{z}}^{\intercal}\, \mathbf{B}_{xy,m}\,{\mathbf{z}} =1}{}{}
\addConstraint{\sum\limits_{m = 1}^{M} {{\mathbf{z}}^{\intercal}\,\mathbf{B}_{\beta, m}\,{\mathbf{z}}}  \le 1}{\label{eq:pb12_1}}{}
\addConstraint{{\mathbf{z}}^{\intercal}\, \mathbf{B}_{\beta x y, m}\,{\mathbf{z}}\leq 0,}{\label{eq:pb12_9}}{}
\end{mini!}
where
\begin{equation*}
\mathbf{B}= \begin{bmatrix*}[l]
\mathbf{0}_{(3M+1)\times (3M+1)}  &\frac{1}{2}\mathbf{e}_{(3M+1)}\\
\frac{1}{2}\mathbf{e}_{(3M+1)}^\intercal & 0
\end{bmatrix*}, 
\end{equation*}
\begin{equation*}
\mathbf{B}_{y,m}=\begin{bmatrix*}[l] \mathbf{A}_{y,m} &-\frac{1}{2}{\mathbf{e}}_{M+m}\\
-\frac{1}{2}{\mathbf{e}}_{M+m}^{\intercal} & {0}\\
\end{bmatrix*}, 
\, \mathbf{b}_{xy,m} = \mathbf{e}_{m}+\mathbf{e}_{m+M}\:,\nonumber
\end{equation*}
\begin{equation*}
 \mathbf{B}_{xy,m}=\begin{bmatrix*}[l] \mathbf{0}_{(3M+1)\times (3M+1) } &\frac{1}{2}{\mathbf{b}}_{xy,m}\\
\frac{1}{2}{\mathbf{b}}_{xy,m}^{\intercal} & {0}\\
\end{bmatrix*}, 
\end{equation*}

\begin{equation*} 
\mathbf{B}_{\beta, m}=\begin{bmatrix*}[l]\mathbf{0}_{(3M+1) \times (3M+1) } & \frac{1}{2} {\mathbf{e}}_{m+2M}\\
\frac{1}{2} {\mathbf{e}}_{m+2M}^{\intercal} &0\\
\end{bmatrix*}, 
\end{equation*}

\begin{equation*}
\mathbf{B}_{\beta x y, m}=\begin{bmatrix*}[l]\mathbf{A}_{\beta x y, m} & \frac{1}{2}{\mathbf{b}}_{cy,m} \\
\frac{1}{2} {\mathbf{b}}_{cy,m}^{\intercal} &0\end{bmatrix*},  
\mathbf{B}_{x,m}= \begin{bmatrix*}[l]
\mathbf{A}_{x,m}  &\frac{1}{2}\mathbf{e}_{m}\\
\frac{1}{2}\mathbf{e}_{m}^\intercal & 0
\end{bmatrix*},
\end{equation*}
\begin{align} 
& \mathbf{A}_{\beta x y, m}=\mathbf{A}_{\beta x,m} + \mathbf{A}_{\beta y,m}+
\mathbf{A}_{\beta t,m} ,\, \mathbf{b}_{cy,m} = k_{m}^{c}\mathbf{e}_{m+M}\:.\nonumber
\end{align}
Next, we apply the SDR to obtain the desired results. Let $\mathbf{Y} = \mathbf{z}\, \mathbf{z}^\intercal$ with \texttt{rank}$(\mathbf{Y}) =1$. Then, the separable semi-definite programming (SDP) problem can be expressed by relaxing problem (10) is as follows
\begin{mini!}[2]
{{\mathbf{Y}}}{\text{Tr}(\mathbf{B}\,{\mathbf{Y}})} {}{}{\label{eq:pb13_0}} 
\addConstraint{\text{Tr} (\mathbf{B}_{x,m} \,\mathbf{Y}) =0}{\label{eq:pb13}}{}
\addConstraint{\text{Tr} (\mathbf{B}_{y,m}\, \mathbf{Y}) =0}{\label{eq:pb13}}{}
\addConstraint{\text{Tr} (\mathbf{B}_{xy,m}\, \mathbf{Y}) =1}{\label{eq:pb13}}{}
\addConstraint{\sum\limits_{m = 1}^{M} {\text{Tr} (\mathbf{B}_{\beta, m} \mathbf{Y})} \le 1}{\label{eq:pb13_1}}{}
\addConstraint{\text{Tr} (\mathbf{B}_{\beta x y, m} \mathbf{Y}) \leq 0 \:.}{\label{eq:pb13_8}}{}
\end{mini!}
We solve the above SDP problem in a polynomial time using a standard SDP software SeDuMi~\cite{cvx}. We obtain the offloading decision $x_{m}\text{ and }y_{m}$ of the original problem (8) from $\mathbf{Y}$. We use randomization method~\cite{Chen2018TMC} to find binary offloading decisions. Accordingly, the probability of task processing at end-user and edge server is given as 
\begin{subequations}\label{prob}
\begin{align}
    P^l_{m} = \dfrac{p^l_{m}}{p^l_{m}(1-p^e_{m})+(1-p^l_{m})\,p^e_{m}}\:,\\
    P^e_{m} = \dfrac{p^e_{m}}{p^l_{m}(1-p^e_{m})+(1-p^l_{m})\,p^e_{m}}\:,
\end{align}
\end{subequations}
where $p^l_{m} = x_{m}$ and $p^e_{m} = y_{m}$. Now, we generate $N$ i.i.d. feasible offloading solutions as $\boldsymbol{\xi}^{(n)} =
[(q^{(n)}_1 )^{\intercal} \ldots (q^{(n)}_M )^{\intercal} ]^{\intercal}$ using the probabilities in~\eqref{prob}, for $n =1, \ldots,N$, as follows
\begin{align}
q_{m}\!=\!& \begin{cases}
{[1, \,0]} & \text{with probability}  P^l_{m}\text{ (at local CPU)\:,} \\
{[0,\,1]} & \text{with probability}  P^e_{m}\text{ (at edge server)}\:.
\end{cases}
\end{align}
Next, we solve the problem (5) for the optimal resource allocation corresponding to offloading decision $\boldsymbol{\xi}^{(n)}$ obtained using (13). Therefore, \eqref{UEdelay} can be rewritten as 
\begin{subequations}
\begin{align}
T^\textsf{L}_i+T^\textsf{E}_i & = k_i^f+\dfrac{k_i^{\eta}}{\beta_i}, \\
T^\textsf{L}_{j}+T^\textsf{E}_{j} & = k_j^f+\dfrac{k_j^{\eta}}{\beta_{j}},
\end{align} \label{2ndcvx0}
\end{subequations}

 \noindent where $k_i^f = ({D_i\,L\,x_{i}})/f_i^l+  (D_i\,L\,y_i)/{f_i^e},\, {k_i^{\eta}} = ({D_i\,y_{i}})/({\eta_i\,C}),\, k_j^f = ({D_{j}L\,x_{j}})/f_j^l
 + (D_j\,L\,y_j)/f_j^e,$ and $ k_j^{\eta} = (D_j y_j)/(\eta_j\,C) $. Hence, our optimization problem becomes
 \begin{mini!}[2]
{\mathbf{r}}{\max \{(T_{i}^\textsf{L}+ T_{i}^\textsf{E}),\,(T_{j}^\textsf{L}+T_{j}^\textsf{E})\}  \; \forall ~ i\in\mathcal{N}_1,\, j\in\mathcal{N}_2  } {}{}{\label{eq:pb22_0}}
\addConstraint{\sum\limits_{i = 1}^{K} {y_i\beta_i} +\sum\limits_{{j} = K+1}^{M} {y_j\beta_{j}} \le 1}{\label{eq:pb22_2}}{}.
\end{mini!}
Now, we take an auxiliary variable $\theta$ as
\begin{align}
    \max\limits_{i\in\mathcal{N}_1,{j}\in\mathcal{N}_2} \big\{(T_{i}^\textsf{L}+ T_{i}^\textsf{E}),\,(T_{j}^\textsf{L}+T_{j}^\textsf{E})\big\} = \theta\:,\label{2ndcvx_1}
\end{align}
and from \eqref{2ndcvx0} and \eqref{2ndcvx_1}, we can write 
\begin{subequations}
\begin{align}
& k_i^f\,{\beta_{i}}+{k_i^{\eta}}-{\beta_{i}}\,\theta \le 0,\\ 
& k_j^f\,{\beta_{j}}+{k_j^{\eta}}-{\beta_{j}}\,\theta \le 0\:.
 \label{eqn-aux-var-t2}
\end{align}
\end{subequations}
Then, the optimization problem becomes
\begin{mini!}[2]
{\mathbf{r}}{\theta} {}{}{\label{eq:pb21_0}}
\addConstraint{\text{(15b), (17a), and (17b)}\:.}{\label{eq:pb11_1}}{}
\end{mini!}
\subsection{Vector-matrix Formation}
We denote ${\mathbf{v}}=[\beta_1,\,\beta_2,\ldots, \beta_K,\,\beta_{K+1},\,\ldots, \beta_M,\, \theta, 1]^{\intercal}$.  Defining a unit vector as ${\mathbf{\hat {u}}}_{p} = [\mathbf{0}_{1 \times ({p}-1)}, 1, \mathbf{0}_{1 \times (M+2-{p})}]^{\intercal}$, the vector-matrix form of problem (18) is written as
\begin{mini!}[2]
{{\mathbf{v}}}{{\mathbf{\hat {u}}}^{\intercal}_{(M+1)}{\mathbf{v}}} {}{}{\label{eq:pb12_0}}
\addConstraint{\sum\limits_{i = 1}^{K} {\mathbf{\mathbf {b}}}^{\intercal}_{yu,i} {\mathbf{v}}+\sum\limits_{j = K+1}^{M} {\mathbf{\mathbf {b}}}^{\intercal}_{yu,j} {\mathbf{v}}\le 1}{\label{eq:pb22_1}}{}
\addConstraint{\mathbf{b}_{kf,i}^{\intercal} {\mathbf{v}} +\mathbf{b}_{k{\eta},i}^\intercal {\mathbf{v}}+ {\mathbf{v}}^{\intercal} A_{\beta \theta, i}{\mathbf{v}}\le 0 }{}
\addConstraint{\mathbf{b}_{kf,j}^{\intercal} {\mathbf{v}} +\mathbf{b}_{k{\eta},j}^\intercal {\mathbf{v}}+ {\mathbf{v}}^{\intercal} A_{\beta \theta, {j}}{\mathbf{v}}\le 0,}{}
\end{mini!}
where
\begin{align}
   & \mathbf{b}_{kf,i} = {k_i^f} \mathbf{\hat {u}}_{i},\, \mathbf{b}_{k{\eta},i} = {k_i^{\eta}} \mathbf{\hat {u}}_{M+2},\,
   \mathbf{b}_{k,{i}}  = \mathbf{b}_{kf,i} +\mathbf{b}_{k{\eta},i},  \nonumber \\ & \mathbf{b}_{yu,i} = y_i \mathbf{\hat{u}}_i,\,
 \mathbf{b}_{kf,j} = {k_j^f} \mathbf{\hat {u}}_{j},\, \mathbf{b}_{k{\eta},j} = {k_j^{\eta}} \mathbf{\hat {u}}_{M+2},\nonumber\\
    & \mathbf{b}_{k,{j}}  = \mathbf{b}_{kf,j} +\mathbf{b}_{k{\eta},j}, ~ \mathbf{b}_{yu,j} = y_j \mathbf{\hat{u}}_j\:,\nonumber
\end{align}
\begin{align}
& \mathbf{A}_{\beta \theta, {i}}=-\dfrac{1}{2}\left[ {\begin{array}{*{20}{c}} \mathbf{0}_{(i-1) \times (M+2)}\\
 \hdashline[2pt/2pt]
 \mathbf{\hat {u}}_{M+1}^{\intercal} \\
 \hdashline[2pt/2pt]
 \mathbf{0}_{(M-i) \times (M+2)}\\ 
 \hdashline[2pt/2pt]
 \mathbf{\hat {u}}_{i}^{\intercal} \\
 \hdashline[2pt/2pt]
 \mathbf{0}_{(1) \times (M+2)}
 \end{array}} \right],\nonumber \\
 & \mathbf{A}_{\beta \theta, {j}}=-\dfrac{1}{2}\left[ {\begin{array}{*{20}{c}} \mathbf{0}_{(j-1) \times (M+2)}\\
 \hdashline[2pt/2pt]
 \mathbf{\hat {u}}_{M+1}^{\intercal} \\
 \hdashline[2pt/2pt]
 \mathbf{0}_{(M-j) \times (M+2)}\\ 
 \hdashline[2pt/2pt]
 \mathbf{\hat {u}}_{j}^{\intercal} \\
 \hdashline[2pt/2pt]
 \mathbf{0}_{(1) \times (M+2)}
 \end{array}} \right]\:.\nonumber
 \end{align}
Let  ${\mathbf{s}}^{\intercal}=[{\mathbf{v}}^{\intercal}\;1]^{\intercal}$, thus the objective function becomes 
\begin{mini!}[2]
{{\mathbf{s}}}{{\mathbf{s}}^{\intercal}\, \mathbf{H}\,{\mathbf{s}}} {}{}{\label{eq:pb12_0}}
\addConstraint{\sum\limits_{i=1}^{K} \mathbf{s}^{\intercal}\, \mathbf{H}_{\beta, i}\,{\mathbf{s}} + \sum\limits_{j=K+1}^{M} \mathbf{s}}^{\intercal}\, \mathbf{H}_{\beta, j}\,{\mathbf{s}  \le 1}{}{}
\addConstraint{{\mathbf{s}}^{\intercal}\, \mathbf{H}_{\beta k \theta, i}\,{\mathbf{s}} \le 0}{}{}
\addConstraint{{\mathbf{s}}^{\intercal}\, \mathbf{H}_{\beta k \theta, {j}}\,{\mathbf{s}} \le 0,}{}{}
\end{mini!}
where
\begin{equation*}
\mathbf{H}= \begin{bmatrix*}[l]
\mathbf{0}_{(M+2)\times (M+2)}  &\frac{1}{2}\mathbf{\hat {u}}_{(M+1)}\\
\frac{1}{2}\mathbf{\hat {u}}_{(M+1)}^\intercal & 0
\end{bmatrix*}, \nonumber 
\end{equation*}
\begin{equation*}
\mathbf{H}_{\beta, i}= \begin{bmatrix*}[l]
\mathbf{0}_{(M+2)\times (M+2)}  &\frac{1}{2} \mathbf{b}_{yu,i}\\
\frac{1}{2} \mathbf{b}_{yu,i}^\intercal & 0
\end{bmatrix*}, \nonumber 
\end{equation*}
\begin{equation*}
\mathbf{H}_{\beta, j}= \begin{bmatrix*}[l]
\mathbf{0}_{(M+2)\times (M+2)}  &\frac{1}{2} \mathbf{b}_{yu,j}\\
\frac{1}{2} \mathbf{b}_{yu,j}^\intercal & 0
\end{bmatrix*}, \nonumber 
\end{equation*}
\begin{equation*}
\mathbf{H}_{\beta k\theta, i}= \begin{bmatrix*}[l]
\mathbf{A}_{\beta \theta, i}  & \dfrac{1}{2}\mathbf{b}_{k,i}\\
\dfrac{1}{2}\mathbf{b}_{k,i}^\intercal & 0
\end{bmatrix*},
\,\mathbf{H}_{\beta k\theta, j}= \begin{bmatrix*}[l]
\mathbf{A}_{\beta \theta, j}  & \dfrac{1}{2}\mathbf{b}_{k,j}\\
\dfrac{1}{2}\mathbf{b}_{k,j}^\intercal & 0
\end{bmatrix*}\:.
\end{equation*}
Further, applying the SDR to obtain the desired results, let $\mathbf{S} = {\mathbf{s}}\, {\mathbf{s}}^\intercal$ such that \texttt{rank}$(\mathbf{S}) =1$. Then, the SDP problem, by relaxing problem (20), can be expressed as
\begin{mini!}[2]
{{\mathbf{S}}}{\text{Tr}(\mathbf{H}\,{\mathbf{S}})} {}{}{\label{eq:pb23_0}} 
\addConstraint{\sum\limits_{i=1}^{K} \text{Tr} (\mathbf{H}_{\beta, i}\,\mathbf{S}) + \sum\limits_{j=K+1}^{M} \text{Tr} (\mathbf{H}_{\beta, {j}}\,\mathbf{S}) \le 1}{\label{eq:pb23}}{}
\addConstraint{{\text{Tr} (\mathbf{H}_{\beta k \theta, i}\,\mathbf{S})} \le 0}{\label{eq:pb23_1}}{}
\addConstraint{{\text{Tr} (\mathbf{H}_{\beta k \theta, {j}}\,\mathbf{S})} \le 0\:.}{\label{eq:pb23_1}}{}
\end{mini!}
We solve the above SDP problem (21) in a polynomial time, denoting $\mathbf{S}$ as the optimal solution of the SDP problem (21). Finally, we obtain the optimal values of $\beta_i$ and $\beta_j$ from $\mathbf{S}$.

\section{Simulation Results}\label{sec:SimulationResult}
 
\begin{figure}[t]
\centering
\includegraphics[width={0.49\textwidth}]{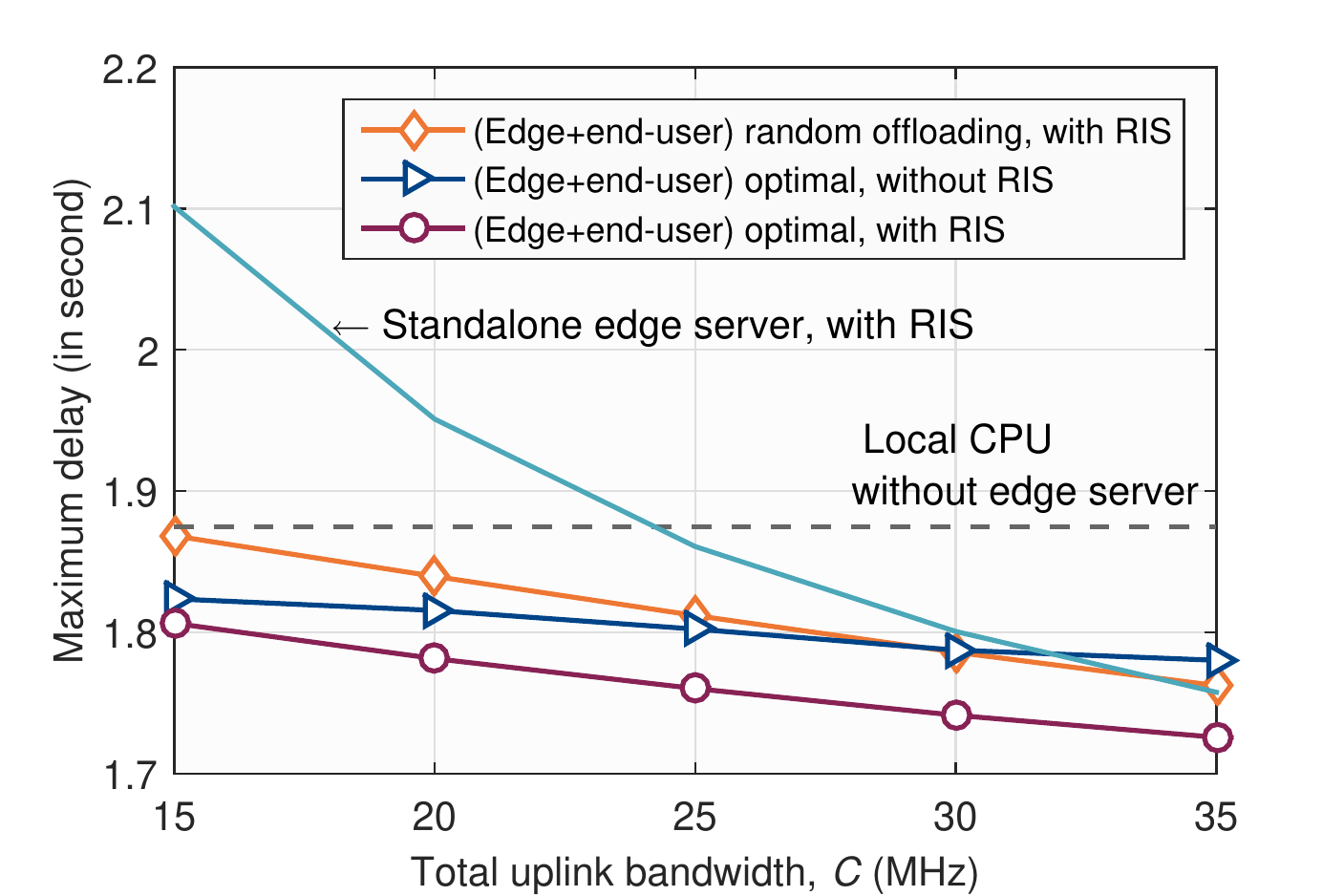}
\vspace{1pt}
\caption{Delay performance vs. total uplink bandwidth.}
\label{fig:Capacity}
\vspace{5pt}
\end{figure}

\begin{figure}[t]
\centering
\includegraphics[width={0.49\textwidth}]{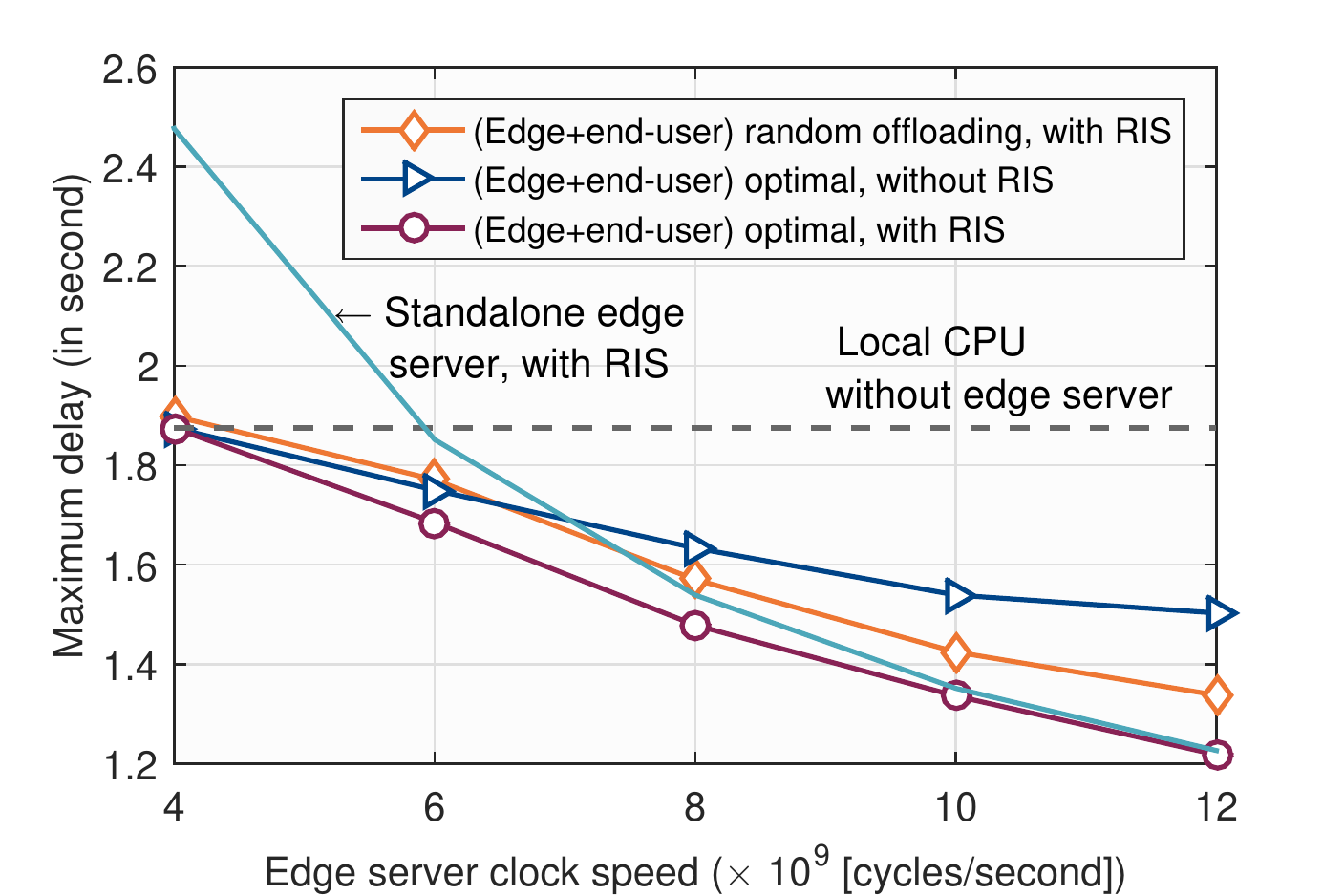}
\vspace{1pt}
\caption{Delay performance with computing resource in edge server.}
\vspace{5pt}
\label{fig:edge_resource}
\end{figure}
In this section, we evaluate the performance of the proposed RIS-assisted computation offloading policy with Monte Carlo simulations. Unless specified, we set CPU clock speed of end-user, i.e, $f_i^l=f_j^l=500\!\times\!10^6$ [cycles/second] and edge server, i.e., $f^\textsf{E}=5\!\times\!10^9$ [cycles/second].
We assume that edge server equally distributes its CPU clock speed, $f^\textsf{E}$, to every end-users, i.e., $f_i^e = f_j^e= f^\textsf{E}/M$. We further consider that the size of a single task is uniformly distributed over $[0.1,0.9]$\,[MB] with a task processing density, $L=1900$ [cycles/byte]. Also, the total available uplink bandwidth is $C=15$ [MHz]. We assume that $M=8$ end-users, out of which $(M-K)=3$ end-users have poor wireless connection. Moreover, we set $\eta_i = 3.5$ [bps/Hz], $\eta_{j} = 0.1$ [bps/Hz] without RIS and $\eta_{j}=3$ [bps/Hz] with RIS. We further set $N=10$ as in~\cite{Chen2018TMC}.  The simulation results\footnote{The source code is available at \url{https://github.com/MithunHub/GC2021Offloading}. } are averaged over at least 10,000 different runs.

Fig.~2 illustrates the maximum delay performance over the network with different uplink bandwidth. When we consider \emph{`standalone edge server'}, the entire data is offloaded to the edge server. Thus, at lower uplink bandwidth, the worst performance is observed due to high uploading delay. From the figure, we can see that the maximum delay decreases with the increase of uplink bandwidth. The main reason is the uplink transmission delay decreases with the increase of uplink bandwidth. Therefore, the maximum delay over the network decreases. It is interesting to see that with RIS assistance, the maximum value of delay further reduces. Note that in this paper, we aim to minimize the maximum delay experienced by any end-user in the networks. Therefore, {when no RIS support is available to the end-users with poor connection, the delay of these end-users has the adverse effect in minimizing the maximum value of delay in the network}. Hence, reducing the uplink transmission delay with RIS assistance results in decreasing the maximum delay over the network. 

Moreover, to show the performance with the computation resources in the edge server, Fig.~3 presents the maximum delay with increasing the CPU cycles/second of the edge server. From the figure, we observe that increase in CPU rate $f^\textsf{E}$ at the edge reduces the maximum delay. Moreover, one can clearly see that the offloading approach with RIS assistance exhibits better performance than the case without RIS assistance. In addition, we observe that at very high CPU cycles/second, the performance of standalone edge server and optimal offloading with RIS assistance gets very close to each others. Because, edge server's CPU rate is so high that end-users always prefer to offload the tasks under the setting of uplink data rate.

\section{Conclusions}\label{sec:Conclusions}
In this paper, we studied computation offloading in a reconfigurable intelligent surface-assisted edge computing system. We employed the benefits of RIS to improve the uploading transmission rate for end-users with poor connection. Our proposed offloading scheme optimized the binary offloading decision variable, the uploading bandwidth allocation, and the CPU frequency allocated for the task data by the edge server. We applied SDR to solve the above QCQP problem. We note that with a better uplink quality, the poor user enjoys more chances to use the computational resources of the edge server, thereby improving the overall network performance than without RIS assistance. Our future work includes studying reliability and deadline constraints in task data offloading for an RIS-assisted edge computing system.

\section*{Acknowledgment}
This work was supported in part by the National Natural Science Foundation of China under Grant 61901128 and Nanjing University of Information Science and Technology Start-up Fund Grant 1521632101005. The corresponding author is Mian Guo.

\bibliographystyle{IEEEtran}


\end{document}